\documentclass[%
 reprint,
 amsmath,amssymb,
 aps,
]{revtex4-1}

\usepackage{graphicx}
\usepackage{dcolumn}
\usepackage{bm}

\begin{document}

\preprint{APS/123-QED}

\title{Structural Relaxation in ‘Simple’ Yield Stress Materials Influences Their Rheology}

\author{Kasra Farain and Daniel Bonn}

\affiliation{Van der Waals–Zeeman Institute, Institute of Physics, University of Amsterdam, Science Park 904,  1098XH Amsterdam, The Netherlands\\
 k.farain@uva.nl\\d.bonn@uva.nl}

\begin{abstract}
Simple yield stress materials are composed of soft particles, bubbles, or droplets with purely repulsive forces. The constituent elements are typically too large to undergo thermal fluctuations, suggesting that the internal structure of the material, and therefore the rheology, should not change over time. We explore the rheology of Carbopol, a prototypical simple yield stress material, and show that gradual structural relaxation of the material results in a small yet significant reduction in the dynamic yield stress. This relaxation process can lead to a non-monotonic creep deformation rate under constant stress, culminating in delayed fluidization of the material. These findings show that the yield stress is not merely a static material property but may be a function of the internal structure of the material.
\end{abstract}

\pacs{Valid PACS appear here}
\maketitle



Yield stress materials behave as viscoelastic solids under low stresses but abruptly yield and begin to flow like liquids once a critical stress threshold is exceeded \cite{1, 2, 3}. This dual solid–liquid nature has made these materials indispensable in industries such as food processing, cosmetics, and pharmaceuticals. Standard protocols for measuring the rheology of yield-stress materials often involve pre-shear treatments to improve reproducibility, even for simple model systems like Carbopol gels, yet reported yield stress values still show considerable variability across studies and laboratories \cite{4, 5, 6, 7, 8}. This raises fundamental questions: Why are such pre-shear procedures necessary to begin with, and how exactly do they influence the response of the material?.

Yield stress materials are generally categorized into two distinct classes based on their time-dependent behavior \cite{6,9,10}. In thixotropic materials, the yield stress increases over time when the material is at rest. During this period, colloidal particles, driven by Brownian motion and interparticle attractions, gradually organize into a more robust, percolated network \cite{1,4, 5, 6,11, 12, 13}. At the threshold stress, when the material yields, this network fragments into smaller components. Under continued shear, the viscosity of the material gradually decreases as additional particle bonds break. As a result, the rheology of thixotropic materials exhibits pronounced dependence on aging and flow history. Achieving steady-state flow can take considerable time under low shear rates and, in some cases, may not be reached within a practical experimental timescale. Within this framework, mechanical disturbances such as shaking or stirring disrupt the internal structure and thereby reduce the yield stress \cite{14}. To recover the original, higher yield stress, the material must be left to rest.

In contrast, simple yield stress materials consist of large (non-Brownian) repulsive particles that are not thought to undergo structural evolution. When subjected to shear, these materials typically reach a steady-state viscosity after a relatively short transient response that often includes a stress overshoot. However, very slow variations in the rheological behavior of the widely used model ‘simple’ yield stress material Carbopol, as well as similar materials, have also been observed over longer timescales and remain a topic of ongoing interest \cite{4,7,8,15,16}. Here, we investigate the time dependence of the yield stress in Carbopol and reveal behaviors that challenge this simple categorization. We show that while Carbopol is not a classical thixotropic material, its yield stress does depend on an internal structure that undergoes slow, spontaneous relaxation. This behavior is fundamentally distinct from thixotropy and, in fact, exhibits the opposite trend.

A remarkable consequence of these relaxation dynamics is the emergence of a non-monotonic deformation rate under constant applied stress. The material can initially resist flow when a stress slightly below the enhanced yield stress is applied. As the yield stress decreases due to internal relaxation, however, the applied stress can eventually exceed the evolving yield stress. This leads to a gradual acceleration of the deformation rate, which culminates in the fluidization of the material after a prolonged delay that can be several hours or more. Similar non-monotonic creep and delayed fluidization, arising from shear banding and other effects, have previously attracted considerable attention \cite{17, 18, 19, 20, 21, 22, 23, 24}. Such phenomena are particularly interesting as ‘simple’ yield stress materials like Carbopol are typically considered athermal systems with internal structures that are stable over time.

We use a standard rheology system, a rough, 50-mm, 1$^\circ$ cone--plate geometry in an Anton Paar MCR 302 rheometer, for the relaxation and rheology experiments reported here. The preparation method of Carbopol (0.5 wt\%, Carbopol Ultrez U10 grade) used in the experiments is described elsewhere \cite{7}. Carbopol consists of crosslinked polyacrylic acid grains that swell in water to form soft, deformable microgel particles \cite{16}. These swollen particles, approximately 2 to 20~$\mu$m in size, jam at sufficiently high concentrations, creating a yield stress material. All experiments reported here were performed at 20~$^\circ$C using a vapor trap to minimize evaporation. It is in fact straightforward to observe structural relaxation in Carbopol. The material is loaded onto the bottom plate (rough sandblasted aluminum) of the rheometer and compressed by the upper cone to fill the gap between the cone and plate, as is usual in rheology experiments. During compression, the excess material flows out of the gap and the normal force measured by the rheometer increases. A significant normal force persists after this compression stops and the surplus material around the cone edge has been carefully removed, often measuring a few Newtons for the 50-mm cone. This elastic force gradually diminishes over time (Fig. 1) without any further imposed deformation as the material undergoes internal structural relaxation.

\begin{figure}
	\begin{center}
		\includegraphics[scale=1]{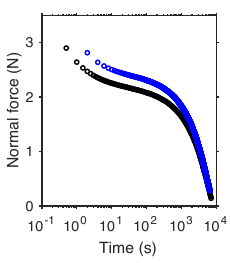}
	\end{center}
	\vspace{-0.5cm}
	\caption{Normal force relaxation in Carbopol under fixed compression without applied shear in a cone--plate rheometer. Data shown for two independently prepared samples. }
	\label{f:numeric}\end{figure}

The steady-state shear stress-shear rate relation for yield stress materials is usually described by the Herschel-Bulkley equation:
\begin{equation}
\Sigma = \Sigma_d + k\dot{\gamma}^n.
\end{equation}
Here, $\Sigma$ is the shear stress and $\dot{\gamma}$ the shear rate. $\Sigma_d$ represents the dynamic yield stress, namely the minimum shear resistance observed at very low shear rates (when $\dot{\gamma} \rightarrow 0$). The parameter $k$ is the consistency index, and $n$ is the shear-thinning exponent \cite{7,15,25, 26, 27, 28}. The Herschel-Bulkley model assumes that the material reaches a steady-state shear stress $\Sigma$ for any given shear rate $\dot{\gamma}$, which may not always be the case. However, if variations in shear resistance occur slowly at very low shear rates ($\dot{\gamma} \rightarrow 0$), the instantaneous value of $\Sigma$ at a given time $t$ approximates the dynamic yield stress ($\Sigma_d$) at that moment. Throughout this work we assume that the shear stress measured following the input of a very low shear rate is effectively the yield stress. 

Figure 2a (left axis) shows the shear stress of a Carbopol sample compressed in the cone-plate geometry and immediately subjected without pre-shear to a very low constant shear rate of 0.02 s$^{-1}$. The material initially experiences a normal force of 1.2 N, which gradually relaxes over time. The shear stress $\Sigma$, which at this low shear rate  roughly corresponds to the yield stress $\Sigma_d$, continues to decrease slowly as the normal stress relaxes, reflecting the structural relaxation in the material. 

The decrease in yield stress during stress relaxation under shear can also be directly observed in stress-controlled experiments. Figure 2b shows the normal force relaxation for a Carbopol sample subjected to a shear rate of 0.02 s$^{-1}$. During the time interval highlighted in red, following 1800 seconds of relaxation, a constant shear stress of 38 Pa is applied. As shown in Fig. 2c (red triangles), the corresponding shear rate exhibits a small peak and then rapidly decays to zero.  This is typical behavior for yield stress materials under applied stresses below the yield stress \cite{18, 20}. The same shear stress of 38 Pa is then applied once again after a further 6900 seconds of relaxation (blue interval). This time the resulting shear rate (Fig. 2c, blue circles) stabilizes at a finite, steady value. This clearly indicates that the yield stress has decreased from above 38 Pa to below 38 Pa between the two tests.

\begin{figure}[hbt]
	\begin{center}
		\includegraphics[scale=1]{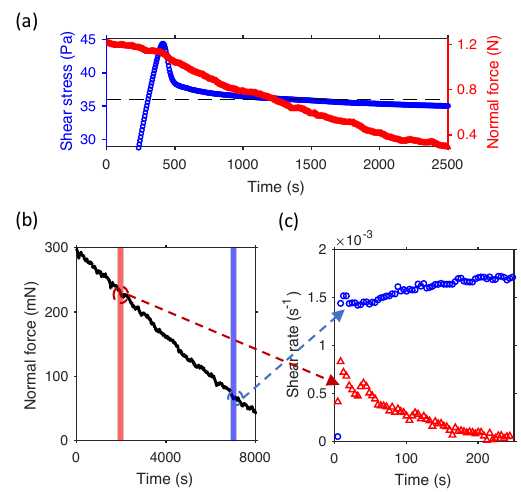}
	\end{center}
	\vspace{-.5cm}
	\caption{(a) Shear stress (left axis) and normal force (right axis) versus time for a Carbopol sample subjected to a constant shear rate of 0.02 s$^{-1}$. The shear stress, which approximates the yield stress at this low shear rate, gradually decreases as the material relaxes. (b) Normal force relaxation for a second Carbopol sample. A constant shear stress of 38 Pa is applied at the time intervals marked by the red and blue bands. (c) Resulting shear rates following the application of 38 Pa. In the first case (red triangles), the shear rate decreases to zero, indicating that 38 Pa is below the yield stress. In the second case (blue circles), after 5400 seconds of relaxation, the same applied stress produces a steady deformation rate, demonstrating that 38 Pa now exceeds the yield stress. }
	\label{f:numeric}\end{figure}

The relaxation of the yield stress of Carbopol is also evident in the shear stress response during up-and-down shear rate sweeps. As shown in Fig. 3a, a freshly loaded Carbopol sample without pre-shear exhibits pronounced hysteresis between the increasing and decreasing shear rate sweeps. This hysteresis arises from structural relaxation that occurs during the measurement: the internal structure evolves during the upward sweep, particularly at high shear rates, such that the material is in a relaxed state during the downward sweep. In Fig. 3b, the shear rate sweep is repeated on the same sample. Because the material has already relaxed during the first experiment, the up and down curves now overlap, indicating a stabilized structure and reduced yield stress.

\begin{figure}[hbt]
	\begin{center}
		\includegraphics[scale=1]{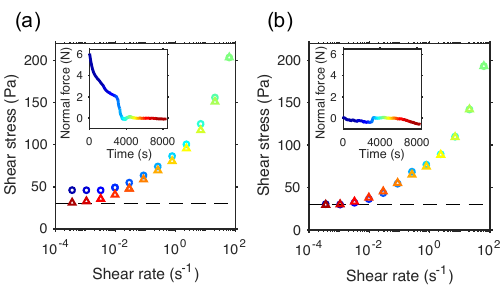}
	\end{center}
	\vspace{-.5cm}
	\caption{Flow curves of Carbopol before and after relaxation. (a) Up-and-down shear rate sweep of a freshly loaded Carbopol sample. The inset shows the relaxation of normal force over time. A clear hysteresis is observed between the upward and downward sweeps, as structural relaxation progresses during the measurement. (b) The shear rate sweep is repeated on the same sample from (a). The up and down sweeps now closely overlap since structural relaxation has been completed.}
	\label{f:numeric}\end{figure}

Thus far, we have demonstrated that both the yield stress and normal stress relax concurrently when a Carbopol sample is first compressed in a cone-plate geometry without pre-shear and subsequently subjected to shear. However, these observations alone do not clarify whether the reduction in yield stress is a secondary effect of the decreasing normal stress or whether the yield stress evolves independently due to structural changes within the material. To address this question we conduct two simple, easily reproducible, yet highly informative experiments.

In the first experiment, we perturb the steady-state flow structure using high-frequency oscillatory shear. The material is first sheared at a constant rate of 0.02 s$^{-1}$ for several hours, allowing it to relax to a steady state with a reduced yield stress. We then apply oscillatory shear at a frequency of 50 Hz and an amplitude of 0.2 radians for 5 minutes. Following this perturbation, the material is again sheared at the same rate of 0.02 s$^{-1}$. As shown in Fig. 4a, the yield stress increases in response to the oscillations and subsequently relaxes slowly over time. Figure 4a (inset) shows the relative increase in yield stress as a function of oscillation amplitude from multiple independent experiments. These results clearly indicate that the reduction in yield stress during shear arises from slow structural changes within the material that are reversible by mechanical agitation.

In the second experiment, we reverse the shear direction from clockwise to counterclockwise after the material has relaxed to its steady-state flow structure and monitor the shear stress response (Fig. 4b). The microstructure necessarily reorganizes with shear reversal. Before reversing the direction, however, we first test the effect of pausing and then resuming shear in the same direction. As shown in Fig. 4b, no significant change in shear stress is observed when shear is stopped at $t = 20,000$ s and restarted 300 s later. In contrast, reversing the shear direction at $t = 30,300$ s results in a pronounced increase in shear stress that takes approximately 2 hours to relax, which is a consequence of the microstructural reorganization induced by the shear reversal. 

\begin{figure}[hbt]
	\begin{center}
		\includegraphics[scale=1]{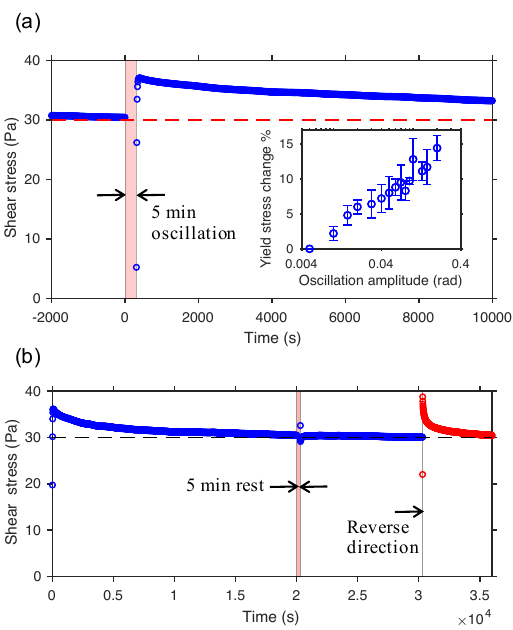}
	\end{center}
	\vspace{-.5cm}
	\caption{Yield stress relaxation is direction-dependent and reversible by mechanical agitation. (a) The material is initially sheared at a constant rate of 0.02 s$^{-1}$ for several hours to reach steady-state conditions. At $t = 0$, oscillatory shear (50 Hz, 0.2 rad amplitude) is applied for 5 minutes. Afterward, the original shear rate is resumed. The oscillatory perturbation induces a sharp increase in yield stress, which gradually relaxes over several hours. The inset shows the relative increase in yield stress as a function of the applied oscillation amplitude from multiple independent measurements. (b) A Carbopol sample is sheared at 0.02 s$^{-1}$ in the clockwise direction. At $t = 20,000$ s, shear is paused for 5 minutes and then resumed in the same direction, resulting in no significant change in yield stress. At $t = 30,300$ s, the shear direction is reversed to counterclockwise, producing a marked increase in yield stress that relaxes over approximately 2 hours.}
	\label{f:numeric}\end{figure}

\begin{figure}[hbt]
	\begin{center}
		\includegraphics[scale=1]{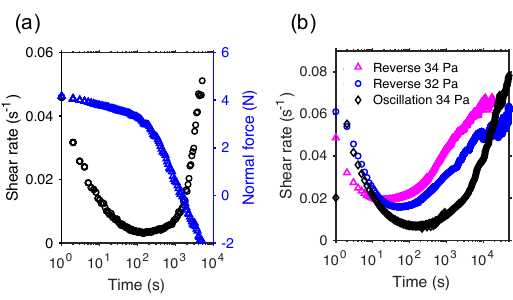}
	\end{center}
	\vspace{-.5cm}
	\caption{Non-monotonic shear rate and delayed yielding in Carbopol under constant stress. (a) A constant shear stress of 42 Pa is applied to a Carbopol sample immediately after loading and compression in the cone-plate system. The normal force (right axis) gradually decreases, indicating stress relaxation. Simultaneously, the shear rate (left axis) initially declines toward zero, suggesting the applied stress is below the yield stress. After approximately 1000 s, however, the shear rate begins to increase, indicating that the yield stress has relaxed below 42 Pa. (b) This non-monotonic shear rate response is reproduced using the oscillatory and reverse shearing protocols from Fig. 4. In the oscillatory case (black curve), the sample first reaches steady-state flow at a shear rate of 0.02 s$^{-1}$, then undergoes oscillatory shear at 50 Hz with a 0.2 rad amplitude for 5 minutes (see Fig. 4a). A constant shear stress of 34 Pa is subsequently applied, and the shear rate is tracked over time. In the reverse shearing case (pink and blue curves), the sample again reaches steady-state flow at 0.02 s$^{-1}$ in the clockwise direction. A constant shear stress of either 32 or 34 Pa is then applied in the counterclockwise direction, and the shear rate is recorded. In all cases, the applied stress is selected to lie above the relaxed yield stress (30 Pa) but below the transiently elevated yield stress induced by oscillatory shear or shear reversal (Fig. 4).}
	\label{f:numeric}\end{figure}

Finally, we demonstrate that yield stress relaxation can result in a non-monotonic strain rate response and delayed yielding in Carbopol under a constant applied stress. Figure 5a shows the shear rate as a function of time for a freshly loaded Carbopol sample without pre-shear under a constant shear stress of 42 Pa. Initially, the shear rate declines toward zero, indicating that the applied stress is below the instantaneous yield stress. As the yield stress gradually relaxes over time and falls below 42 Pa, however, the shear rate begins to increase. This delayed yielding behavior is replicated in Fig. 5b using the oscillatory and reverse shear protocols described in Fig. 4. In these cases, a constant stress is applied that is lower than the transiently elevated yield stress following oscillatory shear (Fig. 4a) or flow reversal (Fig. 4b), but exceeds the relaxed yield stress. As a result, the shear rate initially drops but later increases once the yield stress falls below the applied stress level.

In summary, we have demonstrated that Carbopol, an archetypical simple yield stress material, below the yield point behaves as a viscoelastic solid that undergoes structural relaxation, and that the stress required for fluidization, the yield stress, depends on the structural state. This behavior is fundamentally different from classical thixotropy: here, mechanical perturbations such as oscillatory shear or flow reversal elevate the yield stress rather than reducing it, and the material subsequently relaxes to a lower yield stress state over time. This phenomenology resembles recent observations in colloidal gels, where nanoscale particles aggregate into elastic networks whose mechanical and flow properties vary with ultrasonic vibrations \cite{29,30}. In those systems, ultrasonic vibrations dramatically decrease the gel yield stress and accelerate shear-induced fluidization, with structural and mechanical changes persisting long after vibrations cease.

Our findings have important implications for rheological characterization and application of yield stress fluids. Standard rheological protocols typically employ pre-shearing to eliminate loading-induced structure. Our results demonstrate that structure-dependence is intrinsic to these materials and must be incorporated into their constitutive description. The relaxation dynamics we observe imply that yield stress measurements following pre-shear are protocol-dependent, potentially explaining the variability in reported values across laboratories. Moreover, this structure-dependence becomes critical in contexts where controlled pre-shear is impractical—including industrial processing, storage, and field applications. While pre-shear effects on colloidal gel structure and mechanics have been previously investigated \cite{31}, the spontaneous structural evolution and mechanical rejuvenation we report in Carbopol reveal distinct physics.

Understanding the microscopic origin of this structural relaxation—whether involving particle rearrangements, local packing changes, or contact network modification—requires further investigation through techniques such as confocal microscopy or computational modeling. Such insights are essential for developing predictive constitutive models applicable to industrial and field conditions.

\vspace*{0.3 cm}
This project has received funding from the European Research Council (ERC) under the European Union’s Horizon 2020 research and innovation program (Grant agreement No. 833240). We thank Roos Scheermeijer for her contributions to preliminary experiments and Morton Denn for fruitful discussions.

\end{document}